\documentclass[letterpaper,preprintnumbers,english,floats,floatfix,amssymb,prd,superscriptaddress,twocolumn,nofootinbib]{revtex4-2}
\usepackage{amssymb,amsmath}
\usepackage{epsfig,hyperref}
\usepackage[svgnames]{xcolor}
\usepackage{pgf,tikz}
\usepackage{epstopdf}
\usepackage[british]{babel}
\usepackage{enumerate}
\usepackage{enumitem}
\usepackage{makecell}
\usepackage{booktabs}
\usepackage{graphicx}
\usepackage{nicefrac}
\usepackage{units}
\usepackage{mathrsfs}
\usepackage{placeins} %

\makeatletter
\DeclareRobustCommand*{\bfseries}{%
	\not@math@alphabet\bfseries\mathbf
	\fontseries\bfdefault\selectfont
	\boldmath
}
\makeatother

\def\be{\begin{equation}}
\def\ee{\end{equation}}
\def\beq{\begin{eqnarray}}
\def\eeq{\end{eqnarray}}

\makeatletter

\newcommand{\arXiv}[2][]{\href{http://arxiv.org/abs/#2}{\texttt{arXiv:#2\@ifempty{#1}{}{ [#1]}}}}
\makeatother

\begin{document}
\title{Critical collapse of a self-interacting scalar field in asymptotically anti-de Sitter spacetime}
	
\author{Li-Jie Xin}
\affiliation{School of Physics and Optoelectronics, South China University of Technology, Guangzhou 510641, China}
	
\author{Xiangdong Zhang\textsuperscript{1}}
\email{scxdzhang@scut.edu.cn}
	
\date{\today}
	
\begin{abstract}
We study the critical gravitational collapse of a spherically symmetric massless scalar field in asymptotically anti-de Sitter (AdS) spacetime. The scalar field potential adopted here is inversely proportional to the square of the AdS curvature radius $\ell$, and the system admits a well-known exact static solution. Working in polar coordinates, we first confirm that type II critical collapse occurs for a range of distinct initial configurations when $\ell=8$, where the measured echoing period and critical exponent are in excellent agreement with Choptuik’s classic results. We then fine-tune the initial amplitude of the scalar field for a series of AdS radii $\ell$, performing calculations in both polar coordinates and double null coordinates to cross-validate our results. We find that the form of the potential does not alter the critical behavior of gravitational collapse in any meaningful way: in particular, both the echoing period ($\Delta \approx 3.4$) and critical exponent ($\gamma \approx 0.37$) remain essentially unchanged across all tested values of $\ell$.
\end{abstract}
\maketitle

\section{Introduction\label{sec:introduction}}
The study of gravitational collapse originates from the pioneering work of Oppenheimer and Snyder~\cite{Oppenheimer:1939ue}. This process describes the dynamical contraction of a matter distribution under its own self-gravity, and as the fundamental physical mechanism underlying black hole formation, it remains a core active research area in general relativity to this day~\cite{Singh:1994tb,Ilha:1996tc,Markovic:1999di,Mann:2006yu,Schmitz:2019jct,Piechocki:2020bfo,Shojai:2022pdq,Lewandowski:2022zce,Shi:2024vki,Ou:2025bbv,Bueno:2025gjg}.

Choptuik’s landmark discovery of critical phenomena in gravitational collapse established a foundational new framework for testing the cosmic censorship conjecture~\cite{Choptuik:1992jv}. For a one-parameter family of initial data parameterized by $p$, Choptuik showed that there exists a critical threshold $p^{*}$ that separates two distinct end states of the system: for $p < p^{*}$, the scalar field disperses completely to infinity, while for $p > p^{*}$, gravitational collapse proceeds to form a black hole. Exactly at the critical threshold $p^{*}$, the system converges to a universal critical solution that behaves as an intermediate attractor: any initial configuration sufficiently close to $p^{*}$  will linger near this critical solution before evolving to its final asymptotic state, and the closer $p$ is to $p^{*}$, the longer the system remains in this near-critical regime. A second defining property of critical collapse is the power-law scaling of the black hole mass with the initial parameter:
$M_{\mathrm{BH}}\propto (p-p^{*})^{\gamma}$, where the critical exponent $\gamma$ is a universal constant, independent of the detailed form of the initial configuration. Since Choptuik’s pioneering work, the study of critical phenomena in collapse has expanded to a broad range of matter models and background spacetimes~\cite{Hirschmann:1994du,Choptuik:1996yg,Gundlach:1996je,Hod:1996ar,Brady:1997fj,Neilsen:1998qc,Koike:1999eg,Carr:1999xs,Gundlach:1999cw,Choptuik:1999gh,Husa:2000kr,Bizon:2001ju,Olabarrieta:2001wy,Honda:2001xg,Oren:2003gp,Ventrella:2003fu,Akbarian:2014gna,Jimenez-Vazquez:2022fix,Reid:2023jmr}. More recently, this line of research has been extended to modified gravity theories~\cite{Liebling:1996dx,Eardley:1995ns,Golod:2012yt,Zhang:2016kzg,Hatefi:2019ews,Benitez:2020szx,Benitez:2021zjs,Gambini:2021uzf} and axially symmetric systems~\cite{Abrahams:1993wa,Choptuik:2003as,Choptuik:2003ac,Baumgarte:2019fai,Fernandez:2022hyx,Baumgarte:2023tdh,Baumgarte:2023saw,Marouda:2024epb,Gundlach:2024eds}. 

Hamade and Stewart were among the first to apply the double null formalism to the study of spherically symmetric collapse of massless scalar fields~\cite{Hamade:1995ce}. Their numerical results confirmed that the final fate of the system—either complete dispersal of the scalar field to infinity or gravitational collapse to a black hole—is determined solely by the amplitude of the initial data, and they reproduced Choptuik’s original results for critical collapse with high precision. As a numerical framework for modeling strong-field gravitational dynamics, the double null formalism is widely favored for its ability to naturally resolve event horizons without coordinate pathologies, and for its simplified treatment of asymptotic boundary conditions. It has since become a standard numerical tool for studying open problems including the cosmic censorship conjecture and the mass inflation instability in black hole interiors. Leveraging these unique advantages, numerous subsequent studies have employed the double null formalism to investigate critical gravitational collapse across a wide range of physical setups~\cite{Ayal:1997ab,Hod:1998gy,Thornburg:2010ick,Borkowska:2011zh,Nakonieczna:2012in,Eilon:2015axa,Nakonieczna:2018tih}.

Previous study was mainly carried out in the asymptotically flat setting. Pretorius and Choptuik conducted the first investigation of critical gravitational collapse for a massless scalar field in $(2+1)$-dimensional asymptotically AdS spacetime~\cite{Pretorius:2000yu}. They observed that a spacelike singularity forms inside the event horizon at threshold, and found that near the black hole formation threshold, the near-critical evolution is described by a discretely self-similar solution, with a mass-scaling critical exponent measured to be approximately $1.2$. This work initiated an extensive line of research into gravitational collapse in asymptotically AdS spacetimes~\cite{Bizon:2011gg}, bounded by the standard reflective boundary condition at asymptotic infinity. They discovered that even arbitrarily small initial perturbations undergo repeated reflections off the AdS boundary, leading to turbulent nonlinear energy concentration that ultimately triggers gravitational collapse and black hole formation. This seminal result sparked a wave of new research into the nonlinear dynamics of confined asymptotically AdS spacetimes, which has yielded substantial progress in our understanding of this regime over the past decade~\cite{Maliborski:2012gx,deOliveira:2012dt,Maliborski:2014rma,Lubbe:2014hpa,Craps:2014vaa,Balasubramanian:2014cja,Bizon:2015pfa,Cai:2015jbs,Evnin:2021buq}.

In Ref.~\cite{Martinez:2004nb}, Martinez et al. constructed an exact static hairy black hole solution in asymptotically AdS spacetime, sourced by a self-interacting scalar field with a potential inversely proportional to the square of the AdS curvature radius $\ell$. This solution is a canonical example of an asymptotically AdS hairy black hole: it possesses a constant-curvature negative event horizon that shields the singularity at the origin, and the scalar field remains regular everywhere outside the origin. Since the strength of the scalar potential is explicitly controlled by the AdS radius $\ell$, this model provides an ideal testbed to study how the potential affects the dynamics of critical gravitational collapse, which we explore in this work.

The paper is organized as follows. In Sec.~\ref{sec:methodology}, we detail the methodology for deriving the dynamical equations in two distinct coordinate systems. Secs.~\ref{sec:tr} and \ref{sec:uv} present the numerical results of critical gravitational collapse obtained through each procedure. Finally, a summary and discussion of our main results are provided in Sec.~\ref{sec:summary}. Throughout this paper, geometric units are adopted such that $G=c=1$.

\section{Methodology\label{sec:methodology}}
\subsection{polar coordinates}
We consider the action for a minimally coupled scalar field in the presence of a negative cosmological constant ($\Lambda = -3 \ell^{-2}$). The action is given by 
\be
S = \int d^{4}x \sqrt{-g} \left[\frac{R+6\ell^{-2}}{16\pi }-\frac{1}{2}g^{\mu\nu}\partial_{\mu}\phi\partial_{\nu}\phi-V\left(\phi\right)\right], \label{action}
\ee
where $\ell$ represents the AdS radius. The scalar field potential $V\left(\phi\right)$ takes the form: 
\be
    V\left(\phi\right) = -\frac{3}{4\pi\ell^{2}}\sinh^{2}\sqrt{\frac{4\pi}{3}}\phi.\label{potential}
\ee
As shown in Equation~(\ref{potential}), the potential is inversely proportional to the square of $\ell$.
The equations of motion are derived by varying the action with respect to the metric $g_{\mu\nu}$ and the scalar field $\phi$:
\be
G_{\mu\nu}-\frac{3}{\ell^{2}}g_{\mu\nu}=8\pi T_{\mu\nu},\label{Einsteinequ}
\ee
\be
\Box \phi -\frac{dV}{d\phi}=0,\label{KGequ}
\ee
where $T_{\mu\nu}$ is the energy-momentum tensor for the scalar field, given by 
\be
    T_{\mu\nu} = \partial_{\mu}\phi\partial_{\nu}\phi -\frac{1}{2}g_{\mu\nu}g^{\alpha\beta}\partial_{\alpha}\phi\partial_{\beta}\phi - g_{\mu\nu}V\left(\phi\right).
\ee

We adopt polar coordinates for the metric, which takes the form
\be
ds_{1}^{2} = -A\left(t,r\right)e^{-2\delta\left(t,r\right)}dt^{2}+\frac{1}{A\left(t,r\right)}dr^{2}+r^{2}d\Omega^{2}.\label{line_element1}
\ee
We introduce the auxiliary variables $\Phi=\phi'$ and $\Pi=\dot{\phi} e^{\delta}/A$, where the prime denotes the derivative with respect to $r$ and the overdot denotes the derivative with respect to $t$. The equations of motion then take the form:
\be
A'  = \frac{1-A }{r} - 4\pi A  r \left(\Pi^{2} +\Phi^{2} \right) - 8 \pi r V\left(\phi\right) + \frac{3r}{\ell^{2}},\label{Einstein_tt}
\ee
\be
\delta'=-4\pi r\left(\Pi^{2} +\Phi^{2} \right) ,\label{Einstein_rr}
\ee
\be
\dot{\phi} = A \Pi e^{-\delta},\label{phi_evo}
\ee
\be
\dot{\Pi} = e^{-\delta}\left(\Phi A'+A \Phi'-A\Phi \delta' + \frac{2A\Phi}{r}-\frac{dV\left(\phi\right)}{d\phi}\right).\label{KGeq}
\ee

To ensure regularity at the origin, we impose $A\left(t,0\right)=1$ and $\Phi\left(t,0\right)=0$. Furthermore, we set $\delta\left(t,0\right) = 0$, which identifies $t$ as the proper time at the origin. The initial conditions are specified as follows:
\be
\phi\left(0,r\right) = \epsilon_{1} \exp\left[-\frac{\left(r-r_{0}\right)^{2}}{\sigma_{1}^{2}}\right],\label{initial_phi}
\ee
\be
\Pi\left(0,r\right) = 0.\label{initial_Pi}
\ee

The numerical algorithm proceeds as follows. We first specify the initial profiles for the scalar field variables $\phi$ and $\Pi$. The metric functions $A$ and $\delta$ are then obtained by integrating the constraint Equations (\ref{Einstein_tt}) and (\ref{Einstein_rr}) using the second-order Runge-Kutta method. Subsequently, the scalar field variables $\phi$ and $\Pi$ are advanced to the next time step via a fourth-order Runge-Kutta integration of the evolution Equations (\ref{phi_evo}) and (\ref{KGeq}). With these updated field values, the metric functions at the new time level are computed by integrating the constraint equations. This procedure is iterated to construct the full spacetime evolution of both the metric and the scalar field. Finally, in order to accurately extract the echoing period, a mesh refinement technique is employed~\cite{Zhang:2016kzg}.

To monitor black hole formation, we employ the Misner-Sharp mass $m\left(t,r\right)$, which represents the total energy enclosed within a sphere of radius $r$. For an asymptotically AdS spacetime, which is given by
\be
\begin{split}
m&\equiv\frac{r}{2}(1-g^{\mu\nu}r_{,\mu}r_{,\nu}+\frac{r^{2}}{\ell^{2}})\\
&=\frac{r}{2}\left(1-A+\frac{r^{2}}{\ell^{2}}\right).\\
\end{split}
\label{MS}
\ee
Black hole formation is identified when the ratio $2m/r$ exceeds $0.9$, at which point the numerical evolution is terminated.

\subsection{double null coordinates}
In this subsection, we adopt double null coordinates, where the line element takes the form
\be
ds_{2}^{2} = -a^{2}\left(u,v\right)dudv + r^{2}\left(u,v\right)d\Omega^{2}.\label{line_element2}
\ee
The corresponding equations of motion in these coordinates are 
\be
r_{,uu}+4\pi r \phi^{2}_{,u}-\frac{2r_{,u}a_{,u}}{a} =0,\label{Einstein_uu}
\ee
\be
r_{,vv}+4\pi r \phi^{2}_{,v}-\frac{2r_{,v}a_{,v}}{a} =0,\label{Einstein_vv}
\ee
\be
\frac{r_{,uv}}{r}+\frac{a_{,uv}}{a}-\frac{a_{,u}a_{,v}}{a^{2}}+4\pi \phi_{,u} \phi_{,v}-2\pi a^{2}V\left(\phi\right)+ \frac{3a^{2}}{4\ell^{2}} =0,\label{Einstein_uv}
\ee
\be
r\phi_{,uv}+r_{,u}\phi_{,v}+r_{,v}\phi_{,u}+\frac{ra^{2}}{4}\frac{dV\left(\phi\right)}{d\phi}=0.\label{KG_uv}
\ee
We introduce the following auxiliary variables:
\be
\begin{aligned}
s&=\sqrt{4\pi}\phi,\quad p = s_{,u},\quad q=s_{,v}, \\
c&=\frac{a_{,u}}{a},\quad d=\frac{a_{,v}}{a},\quad f = r_{,u},\quad g = r_{,v}.\label{aux_uv}
\end{aligned}
\ee
Using these variables, the system can be written as a set of first-order differential equations:
\be
f_{,u} - 2fc + rp^{2}=0,\label{evo_f}
\ee
\be
g_{,v} - 2gd + rq^{2}=0,\label{con_g}
\ee
\be
rf_{,v} + fg + \frac{a^{2}}{4}- 2\pi r^{2}a^{2}V\left(\phi\right)=0,\label{con_f}
\ee
\be
rg_{,u} + fg + \frac{a^{2}}{4}- 2\pi r^{2}a^{2}V\left(\phi\right)=0,\label{evo_g}
\ee
\be
r^{2}c_{,v} + r^{2}pq -fg -\frac{a^{2}}{4} =0,\label{con_c}
\ee
\be
r^{2}d_{,u} + r^{2}pq -fg -\frac{a^{2}}{4} =0,\label{evo_d}
\ee
\be
r p_{,v} + fq + gp +\sqrt{4\pi} r \frac{a^{2}}{4}\frac{dV\left(\phi\right)}{d\phi} =0,\label{con_p}
\ee
\be
r q_{,u} + fq + gp +\sqrt{4\pi} r \frac{a^{2}}{4}\frac{dV\left(\phi\right)}{d\phi} =0.\label{evo_q}
\ee

The boundary conditions at the origin $(u=v)$ are given by $r=0$, $f=-g$, $a=2g$, $p=q$, $a_{,r}=0$, and $s_{,r} = 0$. The initial profiles for $d$ and $q$ are specified as follows:
\be
d\left(0,v\right) = \frac{\tan\left(v/2\ell\right)}{2\ell},\label{ini_d}
\ee
\be
q\left(0,v\right) = \left[2\epsilon_{2}v-\frac{2\epsilon_{2}v^{2}}{\sigma_{2}^{2}}\left(v-v_{0}\right)\right]\exp\left[-\frac{\left(v-v_{0}\right)^{2}}{\sigma_{2}^{2}}\right].\label{ini_q}
\ee

The numerical algorithm employed in this subsection proceeds as follows. The initial values of $q$ and $d$ are specified. The fourth-order Runge-Kutta method is then used to integrate Equations (\ref{aux_uv}), (\ref{con_g}), (\ref{con_f}) and (\ref{con_p}), yielding the initial profiles of $a$, $s$, $g$, $f$, and $p$. With the full set of initial data established, a predictor-corrector algorithm is employed to evolve Equations (\ref{evo_q}) and (\ref{evo_d}) to obtain $q$ and $d$ at the next time step. The remaining variables are subsequently obtained by integrating the constraint equations. This procedure is iterated to generate the full spacetime evolution of the system. 

In our numerical simulations, we employ $8000$ spatial grid points with a uniform resolution of $0.00125$. Black hole formation is identified when the metric function $g$ falls below $10^{-4}$, at which point the numerical evolution is terminated. To investigate the power-law scaling of the black hole mass, we adopt the Hawking mass, defined as:
\be
m_{H} = \frac{r}{2}\left(1+\frac{4 r_{,u}r_{,v}}{a^{2}}+\frac{r^{2}}{\ell^{2}}\right).\label{hawkingmass}
\ee
We further introduce the proper time at the origin $(u=v)$, defined as: 
\be
T = \int_{0}^{u}a\left(w,w\right)dw,\label{propertime}
\ee
which is used to extract the echoing period.

\begin{table*}[htb]
	\centering
	\begin{tabular}{c|c|c|c}    
		\hline
		\rule{0pt}{2.0em}Family  & Form of initial data & Echoing period($\Delta$) &  Critical exponent($\gamma$)  \\
		\hline
		\rule{0pt}{2.0em}	
        $(1)$ & $\phi(r) = \epsilon_{1}\exp\left[-\left(r-r_{0}\right)^{2}/\sigma_{1}^{2}\right]$ &$3.474$ & $0.3730$\\
		\hline
		\rule{0pt}{2.0em}	
        $(2) $ & $ \phi(r) = \epsilon_{1}r^{3}\exp\left[-\left(r-r_{0}\right)^{2}/\sigma_{1}^{2}\right]$     &$3.460$ & $0.3784$\\
		\hline
        \rule{0pt}{2.0em}	
        $(3)$   &$\phi(r) = \epsilon_{1}\left(1-\tanh\left[-\left(r-r_{0}\right)^{2}/\sigma_{1}^{2}\right] \right)$     &$3.417$ & $0.3774$\\
        \hline

	\end{tabular}
\caption{Initial data for the one-parameter families considered in Sec.~\ref{sec:tr}. For families (1)-(3), the amplitude is varied in order to locate the critical threshold and determine the corresponding echoing period and critical exponent. (Both the echoing period and critical exponent are presented with four significant figures.)}
	\label{tab:initial_form}
\end{table*}

\begin{figure*}[t!]
	\centering
	\begin{tabular}{cc}
		\includegraphics[width=0.88\textwidth]{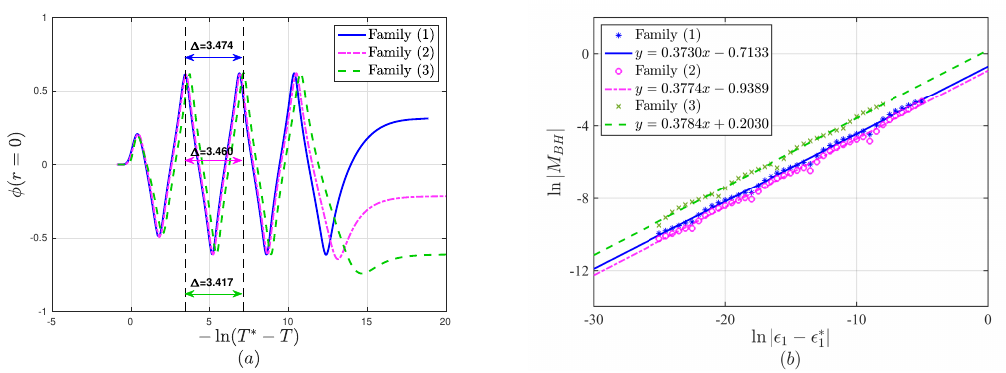}
	\end{tabular}
	\caption{Critical behavior of a scalar field with potential~(\ref{potential}). (a) Evolution of the scalar field $\phi$ at the origin $(r = 0)$ as a function of $-\ln{(T^{*}-T})$, where $T^{*}$ denotes the time of naked singularity formation. (b) Power-law scaling of the black hole mass in the supercritical regime. The AdS radius is set to $\ell=8$. The blue solid line, magenta dash-dotted line, and green dashed line correspond to the Family $(1),(2)$ and $(3)$, respectively.}
	\label{fig:ini}
\end{figure*}

\begin{figure*}[t!]
	\centering
	\begin{tabular}{cc}
		\includegraphics[width=0.88\textwidth]{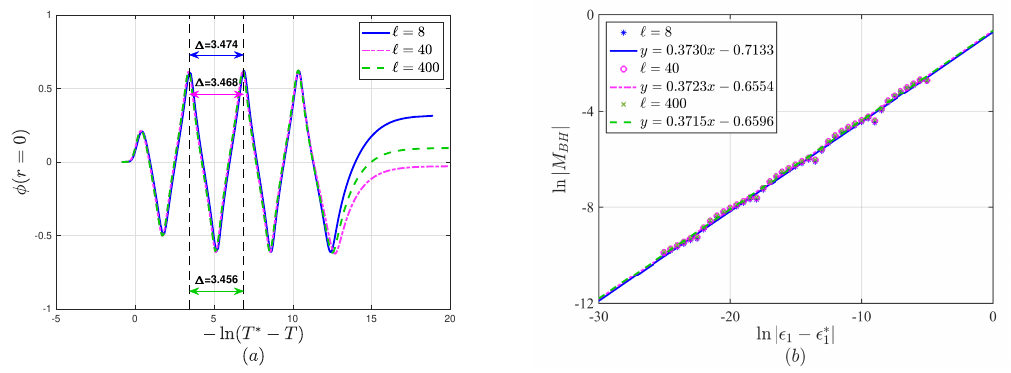}
	\end{tabular}
	\caption{Critical behavior of a scalar field with potential~(\ref{potential}). (a) Evolution of the scalar field $\phi$ at the origin $(r = 0)$ as a function of $-\ln{(T^{*}-T})$, where $T^{*}$ denotes the time of naked singularity formation. (b) Power-law scaling of the black hole mass in the supercritical regime. The initial profile parameters in equation~(\ref{initial_phi}) are $r_{0}=2.0$, and $\sigma_{1}=0.5$. The blue solid line, magenta dash-dotted line, and green dashed line correspond to the cases of $\ell=8,40$ and $400$, respectively.}
	\label{fig:trplot}
\end{figure*}

\begin{table}[htb]
	\centering
	\begin{tabular}{c|c|c}    
		\hline
		\rule{0pt}{1.5em}AdS radius  & Echoing period($\Delta$) &  Critical exponent($\gamma$) \\
		\hline
		\rule{0pt}{1.5em}	$\ell = 8$ &$3.474 $   &$0.3730$  \\
		\hline
		\rule{0pt}{1.5em}	$\ell = 9$  &$3.464 $   &$0.3731 $  \\
		\hline
		\rule{0pt}{1.5em}	$\ell = 10$   &$3.460 $   &$0.3728  $  \\
		\hline
        \rule{0pt}{1.5em}	$\ell = 20$   &$3.441 $   &$0.3719  $  \\
        \hline
        \rule{0pt}{1.5em}	$\ell = 40$   &$3.468 $   &$0.3723  $  \\
        \hline
        \rule{0pt}{1.5em}	$\ell = 60$   &$3.465 $   &$0.3716  $  \\
        \hline
        \rule{0pt}{1.5em}	$\ell = 80$   &$3.489 $   &$0.3724  $  \\
        \hline
        \rule{0pt}{1.5em}	$\ell = 100$   &$3.487 $   &$0.3724  $  \\
        \hline
        \rule{0pt}{1.5em}	$\ell = 200$   &$3.401 $   &$0.3720  $  \\
        \hline
        \rule{0pt}{1.5em}	$\ell = 400$   &$3.456 $   &$0.3715  $  \\
        \hline
        
	\end{tabular}
\caption{The echoing period $\Delta$ (second column), and critical exponent $\gamma$ (third column) for the collapse of scalar fields with different AdS radii $\ell$ are presented. The initial condition in equation~(\ref{initial_phi}) is taken as $r_{0}=2.0$ and $\sigma_{1} = 0.5$. (Both the echoing period and critical exponent are presented with four significant figures.)}
	\label{tab:case1}
\end{table}

\begin{figure*}[t!]
	\centering
	\begin{tabular}{cc}
		\includegraphics[width=0.88\textwidth]{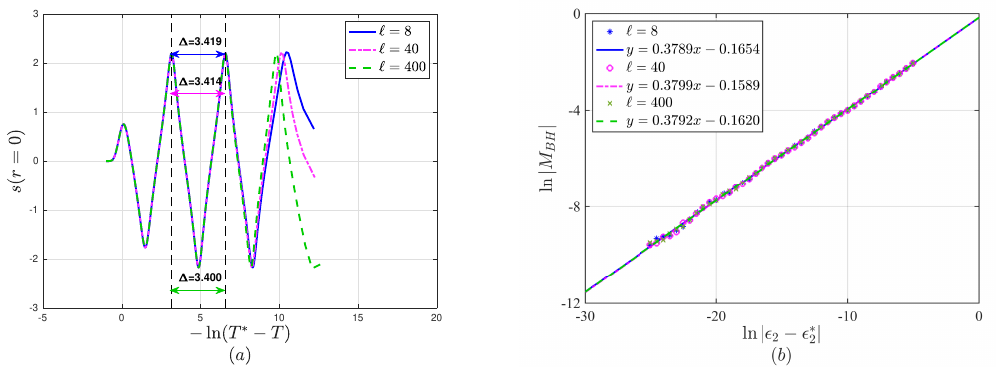}
	\end{tabular}
	\caption{Critical behavior of a scalar field with potential~(\ref{potential}). (a) Evolution of the scalar field $s$ at the origin $(r = 0)$ as a function of $-\ln{(T^{*}-T})$, where $T^{*}$ denotes the time of naked singularity formation. (b) Power-law scaling of the black hole mass in the supercritical regime. The initial profile parameters in Equation~(\ref{ini_q}) are $v_{0}=2.0$, and $\sigma_{2}^{2}=0.5$. The blue solid line, magenta dash-dotted line, and green dashed line correspond to the cases of $\ell=8,40$ and $400$, respectively.}
	\label{fig:uvplot}
\end{figure*}


\section{Result I: polar coordinates\label{sec:tr}}

We begin by numerically simulating a spherically symmetric scalar field with the potential defined in Equation~(\ref{potential}), governed by Equations (\ref{Einstein_tt})-(\ref{KGeq}) in polar coordinates. Type II critical phenomena emerge during the gravitational collapse of an initial scalar field configuration given by equation (\ref{initial_phi}) with an AdS radius $\ell=8$. As shown by the blue solid line in Figure~\ref{fig:ini}, the echoing period and critical exponent are approximately $3.4$ and $0.37$, respectively. We further consider several different initial data profiles at $\ell=8$ to verify that the specific form of the initial configuration does not affect the critical gravitational collapse. These results demonstrate that both the echoing period ($\Delta \approx 3.4$) and the critical exponent ($\gamma \approx 0.37$) are independent of the choice of initial data, as shown in Figure~\ref{fig:ini}. The explicit forms of the initial data and the corresponding numerical results are summarized in Table~\ref{tab:initial_form}, further confirming agreement with Choptuik's results~\cite{Choptuik:1992jv}. 

Throughout the remainder of this section, we adopt the initial data form (1) in Table~\ref{tab:initial_form}, with parameters fixed at $r_{0}=2.0$ and $\sigma_{1}=0.5$. We then search for the critical amplitude $\epsilon_{1}^{*}$, with a precision of $15$ significant digits. The chosen potential depends on the AdS radius $\ell$, and its influence diminishes as $\ell$ increases. We perform numerical simulations for different AdS radii ($\ell=8,40, 400$) and find that critical gravitational collapse occurs in all cases. The resulting echoing periods and critical exponents are in good agreement with previous results, as shown in Figure~\ref{fig:trplot}. Figure~\ref{fig:trplot} (a) shows the evolution of the scalar field at the center as a function of $-\ln{(T^{*}-T)}$, where $T^{*}$ is the time of naked singularity formation. For a scalar field with a potential of the form given in Equation~\ref{potential}, the echoing period remains approximately $3.4$ regardless of the AdS radius $\ell$. These results are consistent with Choptuik’s results~\cite{Choptuik:1992jv}. Figure~\ref{fig:trplot} (b) illustrates the power law scaling of the black hole mass in the supercritical regime for different AdS radii. The results further indicate that the critical exponent remains approximately $0.37$ for different values of the AdS radius, in agreement with Choptuik’s results.   

We further investigate the critical gravitational collapse for ten different AdS radii: $\ell=8,9,10,20,40,60,80,100,200,400$. Numerical simulations for various values of $\ell$ consistently exhibit the characteristic features of type II critical gravitational collapse. In all cases, the echoing period remains around $3.4$, while the critical exponent remains approximately $0.37$. The detailed numerical results are listed in Table~\ref{tab:case1}.



\begin{table}[htb]
	\centering
	\begin{tabular}{c|c|c}    
		\hline
		\rule{0pt}{1.5em}AdS radius  & Echoing period($\Delta$) &  Critical exponent($\gamma$) \\
		\hline
		\rule{0pt}{1.5em}	$\ell = 8$ &$3.419 $   &$0.3789 $  \\
		\hline
		\rule{0pt}{1.5em}	$\ell = 9$  &$3.402 $   &$0.3789 $  \\
		\hline
		\rule{0pt}{1.5em}	$\ell = 10$   &$3.414 $   &$0.3791  $  \\
		\hline
        \rule{0pt}{1.5em}	$\ell = 20$   &$3.424 $   &$0.3787  $  \\
        \hline
        \rule{0pt}{1.5em}	$\ell = 40$   &$3.414 $   &$0.3799  $  \\
        \hline
        \rule{0pt}{1.5em}	$\ell = 60$   &$3.403 $   &$0.3787  $  \\
        \hline
        \rule{0pt}{1.5em}	$\ell = 80$   &$3.404 $   &$0.3795  $  \\
        \hline
        \rule{0pt}{1.5em}	$\ell = 100$   &$3.402 $   &$0.3798  $  \\
        \hline
        \rule{0pt}{1.5em}	$\ell = 200$   &$3.406 $   &$0.3791  $  \\
        \hline
        \rule{0pt}{1.5em}	$\ell = 400$   &$3.400 $   &$0.3792  $  \\
        \hline
        
	\end{tabular}
\caption{The echoing period $\Delta$ (second column), and critical exponent $\gamma$ (third column) for the collapse of scalar fields with different AdS radii $\ell$ are presented. The initial condition in equation~(\ref{ini_q}) is set to $v_{0}=2.0$ and $\sigma_{2}^{2} = 0.5$. (Both the echoing period and critical exponent are presented with four significant figures.)}
	\label{tab:case2}
\end{table}

\section{Result II: double null coordinates\label{sec:uv}}

We perform numerical simulations of critical gravitational collapse for a self-interacting scalar field in double null coordinates, with the scalar potential defined in Equation~(\ref{potential}) and the full system of evolution equations given by Equations (\ref{evo_f})–(\ref{evo_q}). To locate the critical threshold for black hole formation, we employ a bisection search on the initial amplitude of the scalar field, with the initial profile shape fixed by the parameters $v_{0}=2.0$ and $\sigma_{2}^{2} = 0.5$, respectively. The critical amplitude is refined to a precision of $12$ significant digits, ensuring that our near-critical evolutions are sufficiently close to threshold to accurately extract universal scaling parameters. 

For the case $\ell=8$, our simulations yield an echoing period of $\Delta=3.4$ and a critical exponent of $\gamma=0.37$, respectively,  in excellent agreement with the classic results for a free massless scalar field. We then extend our analysis to larger values of the AdS radius, performing fine-tuning simulations for $\ell=40, 400$, As shown in Figure~\ref{fig:uvplot}, the measured echoing period and critical exponent remain essentially unchanged across this range. Further simulations across a wide spectrum of $\ell$ values confirm that the AdS radius has no statistically significant effect on either universal parameter. A complete summary of the 
$\ell$ values tested and the corresponding numerical results for the echoing period $\Delta$ or the critical exponent $\gamma$ is provided in Table~\ref{tab:case2}.

\section{Summary\label{sec:summary}}

In this work, we conduct a systematic numerical investigation of type II critical gravitational collapse for a self-interacting scalar field in asymptotically anti-de Sitter (AdS) spacetime. We work within the ansatz of spherical symmetry, and perform all calculations independently in two distinct coordinate frameworks—standard polar coordinates and double null coordinates—to cross-validate our results and rule out coordinate-dependent artifacts. The scalar field potential we consider is specifically inversely proportional to the square of the AdS curvature radius $\ell$, a form motivated by the existence of a known exact static hairy black hole solution for this model that has been widely studied in the literature.

We begin our analysis in polar coordinates, first fixing the AdS radius to 
$\ell=8$ and evolving a range of different initial scalar field profiles to test the universality of the critical behavior. For every initial configuration we consider, we are able to fine-tune the initial amplitude to $15$ significant digits around the critical threshold for black hole formation, confirming that type II critical behavior occurs in this system for all tested initial data. We extract the echoing period $\Delta$ of the discretely self-similar critical solution and the mass-scaling critical exponent 
$\gamma$ from our near-critical simulations, finding that both quantities are completely insensitive to the choice of initial profile, and their measured values are in excellent agreement with the classic results for a free massless scalar field reported by Choptuik. Extending this analysis to a wide range of different AdS radii 
$\ell$, we find that changing the strength of the scalar potential (which is controlled by 
$\ell$) also has no significant impact on the properties of the critical collapse process. Across all values of 
$\ell$ we study, the echoing period remains centered at $\Delta=3.4$, and the critical exponent remains centered at 
$\gamma=0.37$, with no measurable systematic trend or variation as 
$\ell$ changes. These values match the universal critical parameters for free massless scalar collapse within our numerical error.

To confirm these results are independent of the coordinate system chosen for the simulation, we repeat our fine-tuning experiments for multiple values of 
$\ell$ in double null coordinates, a numerical framework widely known for its advantages in resolving strong-field gravitational dynamics and event horizons. The results obtained in double null coordinates are fully consistent with our findings from polar coordinates: for all tested values of 
$\ell$, the extracted echoing period and critical exponent remain unchanged from their universal values for the massless case, matching Choptuik’s results to within numerical uncertainty.

Taken together, our findings provide strong evidence for the universality of type II critical behavior in scalar field gravitational collapse. We have explicitly demonstrated that the critical properties of the collapse process—the echoing period of the self-similar critical solution and the power-law exponent for black hole mass scaling—are completely insensitive to both the form of the initial scalar profile and the strength of the $\frac{1}{\ell^2}$ scalar potential. This adds to a growing body of work showing that the universal critical exponents for type II collapse depend only on the spacetime symmetry and the matter content, and are not altered by mild scalar self-interactions of the form considered here. Future work will extend this analysis to more general forms of scalar potential and to non-spherical perturbations, to further test the robustness of this universality.

\section*{Acknowledgments}\small
This work is supported by the National Natural Science Foundation of China (NSFC) with Grant No.12275087.

\FloatBarrier	

\end{document}